# Detection of cyclotron resonance using photo-induced thermionic emission at graphene/MoS$_2$ van der Waals interface

Yusai Wakafuji[1], Rai Moriya[1,*], Sabin Park[1], Kei Kinoshita[1], Satoru Masubuchi[1], Kenji Watanabe[2], Takashi Taniguchi[2], and Tomoki Machida[1,*]

[1] *Institute of Industrial Science, University of Tokyo, 4-6-1 Komaba, Meguro, Tokyo 153-8505, Japan*

[2] *National Institute for Materials Science, 1-1 Namiki, Tsukuba 305-0044, Japan*

We demonstrate the detection of cyclotron resonance in graphene by using a photo-induced thermionic emission mechanism at the graphene/MoS$_2$ van der Waals (vdW) Schottky junction. At cyclotron resonance in Landau-quantized graphene, the infrared light is absorbed and an electron–hole pair is generated. When the energy of a photoexcited electron exceeds the band offset energy at the graphene/MoS$_2$ interface, the electron transfer occurs from graphene to the conduction band of MoS$_2$, and the hole remains in graphene. This creates an electron–hole separation at the graphene/MoS$_2$ interface at cyclotron resonance and a photovoltage is generated. The proposed method is an infrared photodetection technique through out-of-plane transport at the vdW junction, which is distinct from the previously reported methods that use in-plane transport in graphene for electronic detection of the cyclotron resonance. Despite the simple structure of our device with a single-vdW junction, our method exhibits a very high sensitivity of ~10$^6$ V/W, which shows an improvement of three orders of magnitude over the previously reported values. Therefore, the **proposed method displays a high potential for cyclotron resonance-based infrared photodetector applications.**

*E-mail: moriyar@iis.u-tokyo.ac.jp; tmachida@iis.u-tokyo.ac.jp

The van der Waals (vdW) heterojunctions formed by various two-dimensional (2D) materials such as graphene, transition metal dichalcogenides (TMDs), and hexagonal boron nitride (*h*-BN) offer a new avenue of heterointerface engineering [1]. Owing to the vdW interactions, we can vertically stack various 2D materials with different band gaps, lattice constants, and electron affinity irrespective of the lattice mismatch and interdiffusion. The band offset and the built-in potential at the interface can be effectively controlled by selecting appropriate combination of 2D materials. This unique property has been utilized to build high performance electronics and optoelectronics devices [2,3,4]. Recently, an electrically tunable Schottky barrier at graphene/semiconducting TMD junction has been of particular interest for its application in high performance electrical switches and sensitive photodetectors in visible wavelength region owing to its simple structure compared with traditional heterostructures [5,6,7,8,9,10]. Previously, our group had demonstrated large current modulation by tuning the Schottky barrier height in graphene/$MoS_2$ [9][11] and graphene/$MoSe_2$ vdW interfaces [12].

In this study, we extend the application of graphene/TMD structure to the far-infrared region by combining its thermionic emission transport with the cyclotron resonance in graphene. As illustrated in Figs. 1(a) and 1(b), a vdW interface between graphene and $MoS_2$ (at the charge neutrality point) exhibits a band offset of ~0.1 eV, and it acts as a Schottky barrier [11,13]. The Fermi level of graphene ($E_F$) changes significantly with the carrier density and it can be raised by electron doping [5,9,11,14]. When $E_F$ is below the conduction band (CB) of $MoS_2$, as depicted in Fig. 1(a), the vdW junction exhibits a high resistance OFF state owing to the presence of a Schottky barrier. When $E_F$ of graphene is raised by electron doping to the same level as the CB of $MoS_2$ (Fig. 1(b)), the Schottky barrier diminishes and the vdW junction exhibits a low resistance ON state. Therefore, the junction acts as a gate-tunable Schottky barrier by controlling the OFF and ON states [5,9,14]. We utilize this mechanism to detect cyclotron resonance in graphene. As illustrated in Fig. 1(c), in graphene/MoS$_2$ heterojunction, the monolayer graphene exhibits Landau quantization in the presence of a magnetic field. Each Landau level (LL) changes its energy according to $\propto \sqrt{B}$ [15,16]. The absorption of light due to the cyclotron resonance occurs when the selection rule $\Delta|N| = \pm 1$ for cyclotron resonance transition is satisfied [17,18,19,20], where $N$ depicts the Landau-level index. The energy of cyclotron resonance transition between $N = 0$ LL to $N = +1$ LL depicted by the solid red arrow in Fig. 1(c) varies from 0 eV to ~0.14 eV in the magnetic field range of 0–12 T, which is close to the band offset energy at graphene/MoS$_2$ junction of ~0.1 eV. In the presence of a high magnetic field, the energy of $N = +1$ LL is higher than the CB of MoS$_2$ as illustrated in Fig. 1(c); therefore, a photoexcited electron from $N = 0$ to $N = +1$ LL under the cyclotron resonance can be transferred to the CB of MoS$_2$ as depicted by the dashed red arrow in Fig. 1(c), and this process is called as photo-induced thermionic emission. In principle, this process causes electron–hole separation at the interface and generates photovoltage. Therefore, the cyclotron resonance of graphene can be detected through the junction photovoltage. In this letter, we demonstrated this concept by fabricating a graphene/MoS$_2$ vdW heterojunction device and investigating its infrared photoresponse.

The schematic of fabricated graphene/ MoS$_2$ vdW heterostructure device consisting of a tri-layer graphene (TLG)/MoS$_2$/monolayer graphene (MLG)/h-BN junction on the SiO$_2$/Si substrate is illustrated in Fig. 2(a). The graphene and hexagonal boron nitride (*h*-BN) flakes as well as nm-thick MoS$_2$ are fabricated by the mechanical exfoliation of natural graphite, bulk $h$-BN crystals, and bulk $MoS_2$ crystals (2D Semiconductors Inc.), respectively. Firstly, a ~30 nm-thick $h$-BN flake is exfoliated on a 290 nm-thick $SiO_2$/highly p-doped-Si substrate. Next, the monolayer graphene (MLG) is exfoliated on a polypropylene carbonate (PPC) film on 290 nm-thick $SiO_2$/Si substrate, which is subsequently transferred over the $h$-BN flake using the dry release transfer method [21]. This MLG serves as the bottom electrode. $MoS_2$ is exfoliated separately on PDMS sheets (Gel-Pak, PF-X4-17mil) and transferred on graphene/$h$-BN structure using the PDMS-based dry transfer method [22]. Finally, TLG is dry transferred on top of the structure by the PPC-based dry transfer method, which is used as the top electrode. An atomic force microscopy (AFM) is used to determine the thickness of the flakes after exfoliation. Using electron-beam (EB) lithography and EB evaporation, we fabricated a 70 nm-thick Pd electrode. For the transport measurements, the sample is mounted in the $^4$He cryostat equipped with a superconducting magnet that applies a magnetic field perpendicular to the sample plane. Using an optical fiber and a light pipe, infrared light from the wavelength tunable $CO_2$ laser ($\lambda$ = ~9.25–10.61 μm, corresponding to the photon energy of $E_{ph}$ = ~0.117–0.134 eV; laser spot size of ~33 mm$^2$ and the power density of $3.77 \times 10^{-2}$ Wcm$^{-2}$ at the sample) is irradiated on the sample. We assume that the irradiated light at the sample is non-polarized as it travels through the optical fiber and the light pipe. Because the top TLG is sufficiently transparent in this wavelength range, the incident light can be absorbed by both, the top TLG and the bottom MLG. The induced photovoltage ($V_{ind}$) in Fig. 2(a) was measured between the top TLG and the bottom MLG using a lock-in amplifier under the modulation of laser light with an optical chopper at the frequency of 11 Hz. For current-voltage (*I-V*) measurements, the lock-in amplifier was replaced by a source measurement unit (Keithley 2400), and the *I-V* curves were measured under the *V*-sweep.

An optical micrograph of the fabricated TLG/MoS$_2$/MLG junction is shown in Fig. 2(b). The thickness of MoS$_2$ was measured to be ~9 layers using AFM. The *I-V* characteristics of the TLG/MoS$_2$/MLG junction is evaluated without light irradiation at ~3K. Fig. 2(c) shows the low-voltage region of the *I-V* curves measured for a different back gate voltage ($V_G$) applied to the Si substrate. The *I-V* curves in Fig. 2(c) exhibit a strong dependence on $V_G$, where the TLG/MoS$_2$/MLG junction offers a high resistance at negative $V_G$ values and a low resistance at positive $V_G$ values, similar to the gate-tunable potential barrier shown in Fig. 1(a) and 1(b). The *I-V* behavior is nearly linear at $V_G$ = +30 and +20 V, which is consistent with the ON state (Fig. 1(b)). With decreasing $V_G$ values (+10, 0, and −10 V), a non-linearity emerges in the *I-V* curves. For the $V_G$ values of −10 V or lower, the junction becomes highly resistive and no obvious current flow is detected in this *V* range as shown in Fig. 2(c), and this corresponds to the OFF state (Fig. 1(a)). A zero bias conductance $G = (dI/dV)_{V=0}$, shown in Fig. 2(d), varies rapidly in the positive $V_G$ range. For $V_G$ > ~15 V, $G$ increases linearly with $V_G$ indicating that the Fermi level of MLG is higher than the bottom of CB of MoS$_2$ in this range; therefore, there is no potential barrier at MLG/MoS$_2$ junction [5,9]. Such a gate-tunable conductance is mainly attributed to the Schottky barrier modulation of the MLG/MoS$_2$ interface [11,13,23]. Here, we can reasonably assume that the gate-dependent transport property of the TLG/MoS$_2$/MLG junction is dominated by the MLG/MoS$_2$ interface. This is owing to the lack of change in the Fermi level of TLG in this region due to the screening of the gate electric field in MLG as well as the large density of states in TLG. This assumption is supported by the fact that the asymmetry in the *I–V* curves in the higher bias voltage range (not shown) is consistent with that of the Schottky barrier at the MLG/MoS$_2$ interface.

We measured the junction photovoltage ($V_{ind}$) of our device under the sweep of $V_G$ and the magnetic field *B* at 3 K, and the results are shown in Fig. 3(a). In this measurement, the wavelength (*λ*) and the energy ($E_{ph}$) of light irradiation are 9.57 μm and 0.130 eV, respectively. In the magnetic field range of 11–12 T, we observed substantially negative $V_{ind}$ peaks located in the $V_G$ range of 5–20 V. No apparent $V_{ind}$ signal is observed at other magnetic field values. The graph of $V_{ind}$ vs. $V_G$ at *B* = 11.0 T is presented in Fig. 3(b). The resonance signal at $V_G$ = 10 V has a much sharper and higher negative peak compared to the signal at $V_G$ = 20 V, and there is no obvious signal in the negative $V_G$ region. The mappings of $V_{ind}$ with varying $V_G$ and *B* values are measured at different wavelengths of laser, and the results are shown in Fig. 3(c). The resonant magnetic field changes with the wavelength implying that the origin of the resonance peak is due to the cyclotron resonance [17,18,20]. For comparison, in Fig. 3(d), we plot the magnetic field dependence of the Landau levels in MLG together with the location of the CB minima of MoS$_2$. The Landau-level energy of MLG can be described using the well-known formula, $E = v_F\sqrt{2e\hbar B}$ [24]; where $v_F$ = 1.1 × 10$^6$ m/s denotes the Fermi velocity of graphene, *e* denotes the elementary charge, *ħ* denotes the reduced Planck constant, and *B* denotes the magnetic field perpendicular to graphene. The photon energy of ~0.130 eV, used here corresponds to the energy associated with the transitions from *N* = −1 to 0 or *N* = 0 to +1 indicated by blue and red arrows, respectively in Fig. 3(d) in the magnetic field range of 11–12 T. As shown in the figure, changing the wavelength of the laser shorter (corresponding to higher laser energy) leads the resonant magnetic field of the cyclotron resonance to be higher;

consistent with the results in Fig. 3(c). In Fig. 3(e), we plot the transition energy between $N = -1$ to $0$ or $N = 0$ to $+1$ as a dashed line and the photon energy $E_{ph}$ dependence of the resonance magnetic field as solid circle. The position of the resonance magnetic field is well described at $v_F = 1.08 \times 10^6$ m/s; this value is typically observed for graphene/$h$-BN structure [25], and it serves as a piece of evidence for the proposed method to detect the cyclotron resonance in MLG through the junction photovoltage measurement at the graphene/MoS$_2$ interface.

The observed cyclotron resonance signal in Fig. 3(a) exhibits asymmetry between positive and negative $V_G$ values. As the charge neutrality point of the MLG is located at ~ 5 V, this corresponds to electron–hole asymmetry in the signal. We have plotted the three lowest Landau levels $N = -2, -1, 0, +1$, and $+2$ of the bottom MLG electrode as dashed lines in Fig. 3(a). The range of $V_G$—where we observed the cyclotron resonance—lies between $N = 0$ and $+1$ LLs, where MLG is electron-doped. As the transition probability for $N = -1$ to $0$ or $N = 0$ to $+1$ transition strongly depends on the Fermi energy $E_F$ of the MLG such that the $N = -1$ to $0$ transition is fully blocked when $E_F$ lies between $N = 0$ and $N = 1$, and vice versa, the measurement results show that the cyclotron resonance signal is only observed for the transition from $N = 0$ to $+1$ LLs even under non-polarized light irradiation. This presents a strong contrast from the previous studies that used in-plane transport of graphene to detect the cyclotron resonance [17,19,20]. For in-plane measurements, all of detection mechanism reported so far is sensitive to both $N = -1$ to $0$ and $N = 0$ to $+1$ transitions; as a result, cyclotron resonance signal appears on both electron and hole-side. Therefore, we think that the electron–hole asymmetry observed in this study originates from the vertical transport through the graphene/MoS$_2$ junction.

To explain the observed asymmetry in the cyclotron resonance signal, we illustrate a simple model for photovoltaic mechanism in Fig. 3(f). In this figure, we represent the cyclotron resonance transitions between $N = -1$ to 0 and $N = 0$ to +1 LLs as the blue and red solid arrows, respectively. As seen in Fig. 3(f), a photoexcitation due to the $N = -1$ to 0 transition generates an electron at the $N = 0$ LL. This excited electron cannot transfer to the CB of $MoS_2$ as depicted by the dashed blue arrow in the figure owing to the absence of available density of states in $MoS_2$. However, the photoexcited electron due to the $N = 0$ to +1 transition is generated at the $N = +1$ LL and this electron can transfer to the CB of $MoS_2$ before its relaxation owing to the built-in electric field at the interface (this is depicted by the dashed red arrow in the figure). Therefore, only the $N = 0$ to +1 transition causes an electron–hole separation at the junction and induces photovoltage. Since both photo-excited electron transfer time from graphene to TMD and carrier relaxation time between LLs in graphene is known to exhibit within similar time scale [26,27,28,29], we think it is possible that there is a photo-excited electron transfer from graphene to $MoS_2$ in our device. We expect a negative photovoltage in this mechanism, which is also consistent with our results. We note that the cyclotron resonance-induced signal shown in Fig. 3(a) has double dip structure such that two different dips are located at around $N = 0$ and $N = +1$ LLs, and signal is significantly suppressed in the middle of these LLs. Because the graphene is in the quantum Hall state in the middle of $N = 0$ and $N = +1$ LLs, we speculate that the photovoltage is significantly reduced in this regime due to the lack of available density of states in graphene electrode. This introduce double dip structure observed in the Fig. 3(a).

In addition, our measurements (Fig. 3(a)) do not show cyclotron resonance signals at lower magnetic fields that are typically observed for the cyclotron resonance in the in-plane graphene devices [20]. This behavior can be explained using the energy diagram of the Landau level shown in Fig. 3(d). In addition to the transitions from $N = 0$ to $+1$ and $-1$ to 0 LLs (red and blue arrows) in high magnetic field, using the wavelength of the laser, in principle $N = -1$ to 2 and $-2$ to 1 LLs (green arrows), and $N = -2$ to 3 and $-3$ to 2 LLs (purple arrows) can be also excited in the lower magnetic fields. However, the energy of the excited electrons produced owing to the lower magnetic field transitions is lower than the CB of the $MoS_2$ as shown in Fig. 3(d); therefore, these electrons cannot generate photovoltage through the photo-induced thermionic emission mechanism. This is another piece of evidence of the photo-induced thermionic emission demonstrated in the graphene/$MoS_2$ junction in this work.

In previous studies, the detection of the cyclotron resonance photovoltage in the in-plane graphene devices is mainly attributed to the photo-thermoelectric effect, which shows sensitivity in the range of 1–1000 V/W [17,18,19,20,30] due to the limitation imposed by the thermopower of graphene. To calculate the sensitivity of our device, we first divided the power density of the laser light at the sample position by the junction area of ~1.4 μm² to estimate the total irradiation of 540 pW; then, we divided the peak height of the photovoltage signal $V_{ind} = 0.37$ mV (Fig. 3(b)) by the total irradiation (540 pW) to obtain the sensitivity of $1.4 \times 10^6$ V/W. Hence, we demonstrate an enhancement in the photovoltage signal by three orders of magnitude in our experiment. Such a high sensitivity cannot be explained with photo-thermoelectric effect; thus, this also support our conclusion that the observed photovoltage is due to the photo-induced thermionic emission at the MLG/$MoS_2$ vdW interface.

In summary, we have demonstrated the photovoltaic detection of the cyclotron resonance of graphene in the graphene/MoS$_2$ vdW junction utilizing the band offset between graphene and MoS$_2$. This scheme can be extended to other wavelength ranges of photodetection by selecting an appropriate combination of band gap as well as band offset at the heterointerface. In addition, the mechanism of photodetection with vertical transport generates a photocurrent that is, in principle, proportional to the junction area.

## Acknowledgements

This work was supported by CREST, Japan Science and Technology Agency under Grant Number JPMJCR15F3; and JSPS KAKENHI Grant Numbers JP19H02542 and JP19H01820.

**Figure captions**

Figure 1

(a,b) Illustration of the band alignment between monolayer graphene (MLG) and MoS$_2$ at zero magnetic field: (a) OFF state when the Fermi level of MLG is lower than the bottom of the conduction band (CB) of MoS$_2$, (b) ON state when the Fermi level of MLG is equal or higher than the bottom of CB of MoS$_2$. (c) Schematic of band alignment at the graphene/MoS$_2$ junction in a magnetic field. The Landau levels of graphene with Landau-level index $N = -2, -1, 0, +1$, and $+2$ are depicted. The red arrows in the figure depicts electron photoexcitation from $N = 0$ to $N = 1$ (solid red arrow) and transfer of the excited electron from MLG to MoS$_2$ (dashed red arrow).

Figure 2

(a) Illustration of the fabricated device structure of tri-layer graphene (TLG)/MoS$_2$/monolayer graphene (MLG) junction. (b) Optical micrograph of the fabricated device. (c) Current-voltage (*I-V*) characteristics of the junction measured at different $V_G$ values at $T = 3$ K. (d) $V_G$ dependence of the zero-bias conductance $G = (dI/dV)$ measured at zero magnetic field and at the temperature of 3.0 K.

Figure 3

(a) Image plot of photo-induced voltage $V_{ind}$ with respect to the back gate voltage $V_G$ and magnetic field *B* measured at $T = 3.0$ K. Here, the wavelength ($\lambda$) and energy ($E_{ph}$) of the irradiated light are 9.57 µm and 0.130 eV, respectively. (b) $V_G$ dependence of $V_{ind}$ at $B = 11.0$ T and $T = 3.0$ K. (c) $V_{ind}$ signals plotted as a function of $V_G$ and *B* measured at different laser wavelengths $\lambda$. (d) Illustration of the change of Landau-level energies with magnetic field. Cyclotron resonance transition from $N = 0$ to $+1$ LL (red solid arrow), $N = -1$ to $0$ LL (blue solid arrow), $N = -1$ to $+2$ ($-2$ to $+1$) LL (green arrows), and $N = -2$ to $+3$ ($-3$ to $+2$) LL (purple arrows) are depicted. The band offset between the charge neutrality point of graphene and the bottom of CB of $MoS_2$ is also depicted. (e) Photon energy of light excitation versus $B$, where the dashed line shows the magnetic field dependence of the cyclotron resonance energy for $N = 0$ to $+1$ transition. (f) Schematic of the band alignment at the graphene/$MoS_2$ junction upon infrared light irradiation. The red and blue arrows depict the cyclotron resonance transitions from $N = 0$ to $+1$ and $N = -1$ to $0$ LLs, respectively.

Figure 1

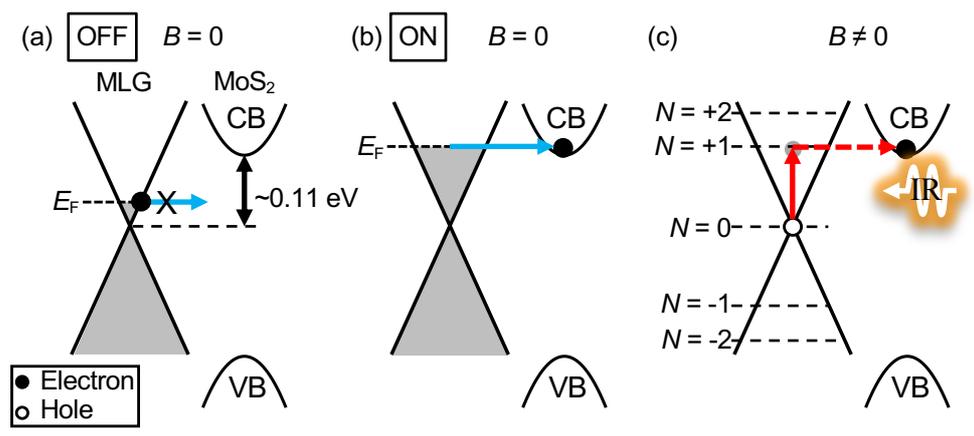

Figure 2

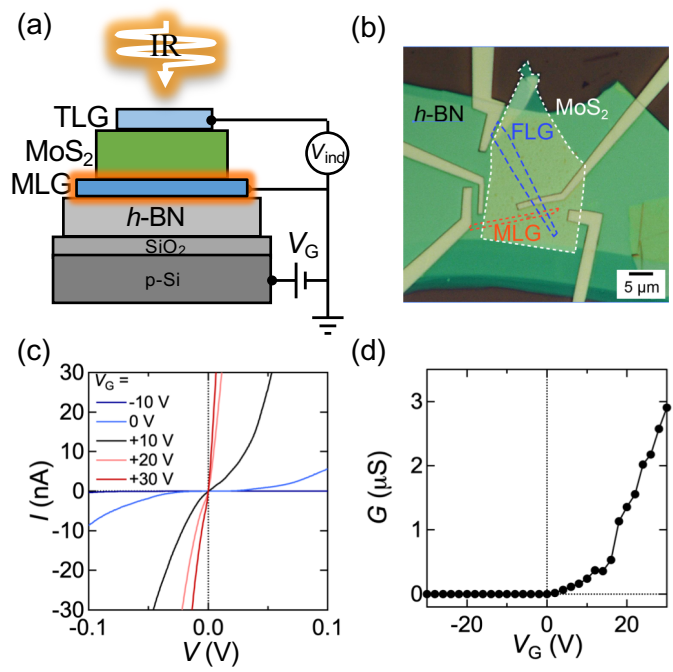

Figure 3

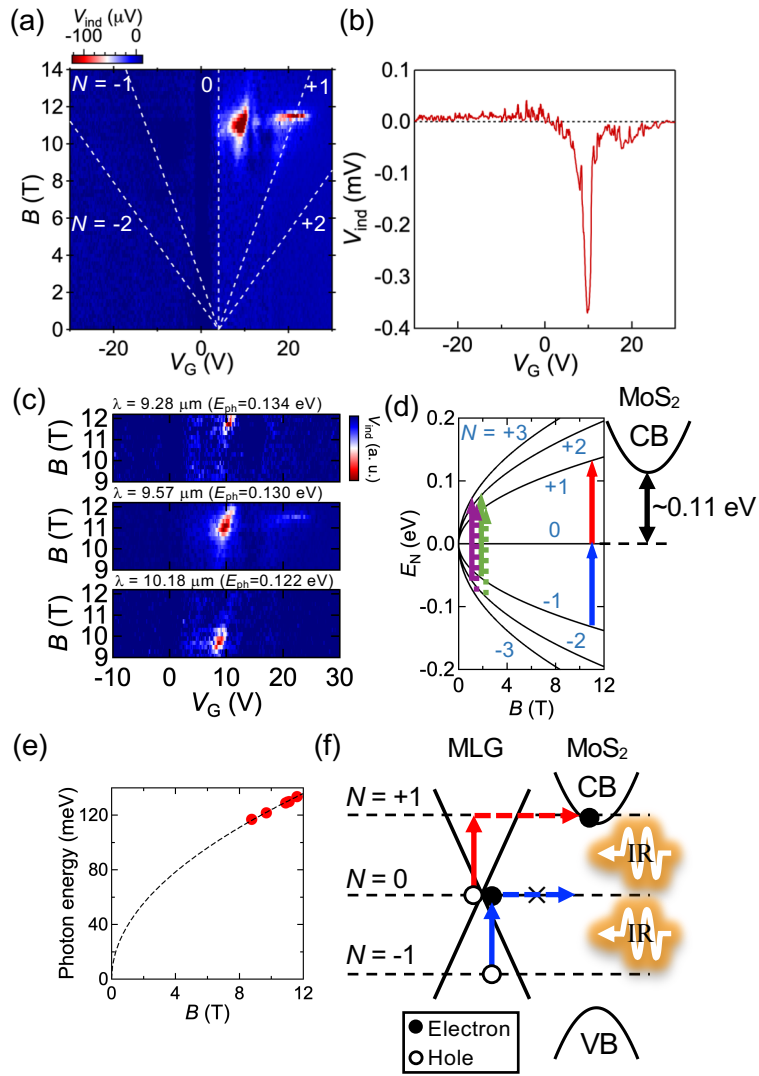